\documentclass[reprint,amsmath,amssymb]{revtex4-2}
\usepackage{float}
\usepackage{graphicx}
\usepackage{dcolumn}
\usepackage{bm}
\usepackage[dvipsnames]{xcolor}
\usepackage{physics}
\newcommand{\red}[1]{{ \color{red} #1}}

\newcommand{\B}[1]{\textbf{#1}}

\begin{document}

\preprint{APS/123-QED}

\title{Symplectic Effective Field Theory for Nuclear Structure Studies}

\author{D. Kekejian}
\email[]{dkekejian@lsu.edu}
\author{J. P. Draayer}
\email[]{draayer@lsu.edu}
\affiliation{Department of Physics and Astronomy, Louisiana State University, Baton Rouge 70803, LA, USA.}
\author{V. I. Mokeev}
\email[]{mokeev@jlab.org}
\affiliation{Thomas Jefferson National Accelerator Facility, 12000 Jefferson Avenue, Newport News, VA, USA.}
\author{C. D. Roberts}
\email[]{cdroberts@nju.edu.cn}
\affiliation{School of Physics, Nanjing University, Nanjing, Jiangsu 210093, China}
\affiliation{Institute for Nonperturbative Physics, Nanjing University, Nanjing, Jiangsu 210093, China}

\date{\today}

\begin{abstract}
A Symplectic Effective Field Theory that unveils the observed emergence of symplectic symmetry in atomic nuclei is advanced.  Specifically, starting from a simple extension of the harmonic-oscillator Lagrangian, an effective field theory applied against symplectic basis states is shown to yield a Hamiltonian system with one fitted parameter.  The scale of the system can be determined self consistently as the ratio of the average volume of a nucleus assumed to be spherical to its volume as determined by the average number of oscillator quanta, which is stretched by the fact that the plane-wave solution satisfies the equations of motion at every order without the need for perturbative corrections. As an application of the theory, results for ${}^{20}$Ne, ${}^{22}$Ne and ${}^{22}$Mg are presented that yield energy spectra, B(E2) values, and matter radii in good agreement with experimentally measured results.
\end{abstract}

\maketitle


\section{Introduction}

Symmetries play a pivotal role in our understanding of interactions that dominate nuclear structure. Their importance extends from quantum field theories such as the SU(3) color group used to proffer an understanding of the quark and gluon dynamics in individual nucleons, to effective field theories and the breaking of chiral symmetry which imposes a critical constraint that generates and controls the dynamics of nuclei in the low-energy regime that can be used to establish a connection with QCD \cite{MACHLEIDT20111, ROBERTS2000S1, CLOET20141, PhysRevLett.106.162002, PhysRevD.91.054035, Eichmann:2016yit}.

Similarly, Effective Field Theories (EFT) have proven to be useful in gaining model-independent approaches in various other analyses \cite{WeinbergEFT,GEORGI1990447,PhysRevLett.114.052501,doi:10.1146/annurev.nucl.52.050102.090637,RevModPhys.81.1773,PAPENBROCK201136}.  The main foundation of any EFT is its ability to exploit a separation of scales between two phenomena, those of interest like low-energy collective nuclear excitations such as rotations and vibrations, and others that focus on the higher energy aspects of the interaction. Notable nuclear collective models\cite{BohrMottelson,Iachello} have been identified as leading-order Hamiltonians of an EFT approach.

Significant progress achieved in the past decade using continuum Schwinger function methods (CSM) paves a way to observe how an effective strong interaction relevant for low-energy nuclear physics emerges from quark and gluon interactions in the strong QCD regime, characterized by a coupling $\alpha_s/\pi \simeq 1$ \cite{Proceedings:2020fyd, Barabanov:2020jvn, Chen:2020wuq, Chen:2021guo}. Furthermore, the CSM results have been rigorously checked in comparisons with a broad array of different experimental results on meson and baryon structure \cite{Carman:2020qmb, Mokeev:2022xfo, Barabanov:2020jvn, Qin:2020rad, Roberts:2021nhw}.

The importance of symmetries is not limited to EFT analyses. For example, it is well-known that SU(3) is the symmetry group of the spherical harmonic oscillator that underpins the many-particle microscopic nuclear shell model \cite{mayer1955elementary} patterned in large part after the atomic case that treats the nucleus as a closed core system of interacting single particles in valence shells with residual interactions. The latter successfully describes single-particle separation energies and binding energies at shell closures called magic numbers; however, it fails to describe effects due to the collective motion of the core such as the emergence of rotational bands in heavy nuclei that can be described phenomenologically by the Bohr-Mottelson collective model \cite{BohrMottelson}, and the fact that the first excited state of the doubly closed shell nucleus of ${}^{16}$O is part of a strongly deformed rotational band that leads to an experimentally observed non-zero quadrupole moment for its ground state.

The SU(3) model advanced by Elliott \cite{Elliot} was the ﬁrst group-theoretical model that captured rotational collectivity in a shell-model framework. One can ﬁnd its roots in the Nilsson model \cite{Nilsson}, which is simply a deformed version of the single-particle shell model. This unveiling of the microscopic origin of collectivity within a nuclear shell-model framework through an algebraic model and the fact that most nuclei are deformed, along with the coexistence of low-lying states in a single nucleus with diﬀerent quadrupole moments \cite{RevModPhys.83.1467}, paved the way to the development of the Sp(3,\B{R}) Symplectic Model \cite{PhysRevLett.38.10}.

The Sp(3,\B{R}) model is a multi-shell extension of the SU(3) model that allows one to organize the spatial parts of many-particle nuclear configurations into a collection of Pauli-allowed shapes. This is a logical first-step of a far more robust theory for grouping many-nucleon configurations into cluster-like shell-model configurations on the lowest rung of what is now known to be an algebraically defined pyramid of deformed eigensolutions coupled through enhanced B(E2) linkages.
Multiple phenomenological and schematic interactions that employ the symplectic symmetry group have been found to give energy spectra, B(E2) quadrupole transitions and radii that are in remarkable agreement with experimental data across the nuclear chart from the lightest systems like ${}^{16}$O  \cite{PhysRevLett.97.202501} and ${}^{12}$C  \cite{DREYFUSS2013511} through to intermediate-mass nuclei spanning from ${}^{20}$Ne \cite{DRAAYER1984215,PhysRevC.89.034312} and ${}^{24}$Mg \cite{ROSENSTEEL19841,CASTANOS1989349,PhysRevLett.82.5221}, up to strongly deformed nuclei of the rare-earth and actinide regions like ${}^{166}$Er \cite{BAHRI2000125} and even ${}^{238}$U\cite{CASTANOS1991469}. 

While such applications of the symplectic model reproduce observed collective patterns in nuclei, they typically rely on schematic or phenomenological interactions. However, recent results from the {\it{ab initio}} Symmetry-adapted No-Core Shell Model (SA-NCSM) \cite{LAUNEY2016101,PhysRevLett.98.162503,PhysRevLett.111.252501,PhysRevLett.124.042501}, that employs realistic chiral effective field theory interactions strongly suggest that the symplectic symmetry is a natural symmetry for nuclei and that its origins should be investigated starting from first principles; that is, from a symplectic effective field theory. Below we show the construction of a symplectic effective field theory, one which when applied to symplectic basis states yields a polynomial of quantum mechanical Hamiltonians for nuclear structure applications. As an application of the theory, results for the ${}^{20}$Ne, ${}^{22}$Ne and ${}^{22}$Mg isotopes are presented.


\section{Symplectic Effective Field Theory}

In this section we present a step-by-step method for building the symplectic effective field theory referenced above. The main concept is to formulate a self-interacting real scalar effective field theory that represents the excitations of a system of $\mathcal{A}$ interacting nucleons. The degrees of freedom of the system are the real scalar fields that represent harmonic oscillator excitations. At leading order, the fields are plane waves that satisfy the equations of motion. For every next-to-leading order, the fields can be taken to be plane waves without the need for perturbative corrections if one imposes a specific requirement on the coupling coefficient of the theory. That requirement, together with the scalar field constraint, set the overall scale of the theory, which can be stated simply in terms of the ratio of the average spherical volume of a nucleus in its ground state to its average volume determined in terms of the number of harmonic oscillator excitations it hosts that allows the system to stretch in ways that are consistent with the pervasive plane-wave constraint.

The construction of this EFT is done through the following steps: In the first subsection (A) we introduce the harmonic oscillator Lagrangian, extend it to $n$-th order, and present the corresponding solutions. In the second subsection (B) we review features of the symplectic Sp(3,\B{R}) group, its generators, and the nomenclature we will use in defining actions of these generators on states within an irreducible representation (irrep) of the symplectic group, especially on the irrep's lowest-weight state from which all others can be built.  In subsection (C) we move to the more familiar Hamiltonian rendering of the dynamics, the details of which -- being quite expansive -- are relegated to an appendix. In subsection (D), the physical features associated with the diagonal elements of the Hamiltonian are examined; while in subsection (E), we do the same for the various off-diagonal elements resulting into a quantum mechanical Hamiltonian applicable for nuclear systems. 

Throughout this paper we will use the Einstein notation for repeated indices and natural units in derivations ($\hbar=c=1$), the following four vector notation for our covariant position vector $x^{\mu}(t,\B{r})$ and $x_{\mu}(t,-\B{r})$ for its contravariant component, $k^{\mu}(E,\B{k})$ for the momentum four vector and $\partial^{\mu}(\frac{\partial}{\partial t},\frac{\partial}{\partial\B{r}})$ for the derivative.  For overall simplicity, we will use $L$ and $H$ for the regular Lagrangian and Hamiltonian, and $\mathcal{L}, \mathcal{H}$ for their density-dependent equivalents, respectively.

\subsection{The Harmonic Oscillator (HO) Lagrangian and its $n$-th order extension}
The simplest Lagrangian density one can write for a real scalar field $\varphi$ is 
\begin{equation}\label{HOLagrangian}
\mathcal{L}=\frac{1}{2}\partial_{\mu}\varphi\partial^{\mu}\varphi,
\end{equation}
which is the Lagrangian density of a harmonic oscillator (HO) for massless excitations (bosons). The classical, still not quantized, fields that satisfy the equations of motion of this Lagrangian are given by the plane-wave solution.
\begin{equation}\label{PW}
\varphi(\B{r},t)=\frac{1}{(2\pi)^{3/2}}\int_{-\infty}^{+\infty}\psi(\B{k},E)e^{ik^{\mu}x_{\mu}}dEd\B{k}.
\end{equation}
The integration is over four variables, the three momenta (\B{k}, a spatial vector) and the energy (E, a scalar). The construction of the effective field theory is accomplished by taking the Lagrangian in Eq.\,\eqref{HOLagrangian} and extending it naturally to its $n$-th order and adding a mass term at every order,
\begin{equation}\label{EFTLagrangian}
\mathcal{L}^{(n)}=\frac{\alpha^n}{2^{n+1}(n+1)!}\big(\partial_{\mu}\varphi\partial^{\mu}\varphi-nm^2\varphi^2\big)^{n+1}\red{.}
\end{equation}
The total Lagrangian density is $\mathcal{L}=\sum_n\mathcal{L}^{(n)}$, where for the $n=0$ term we recover Eq.\,\eqref{HOLagrangian}.
We have added a term $nm^2\varphi^2$, often called ``the mass'' term, at every order. This is added to capture all possible combinations of interaction terms that could result from lowest powers of $\varphi$ and $\partial_{\mu}\varphi\partial^{\mu}\varphi$. Including $\varphi$ only shifts the equations of motion by a constant, therefore the lowest possible power is $\varphi^2$. In this formulation $\alpha$ is the coupling coefficient of the theory. Since the dimension of a Lagrangian density always has to be $[\mathcal{L}]=4$, it follows logically that $[\alpha]=-4$ and as well that this is a \emph{non-renormalizable} EFT. 
The main advantage of this systematic construction of the Lagrangian density given by Eq.\,\eqref{EFTLagrangian} is that it is unique in the sense that the plane wave solution given by Eq.\,\eqref{PW} satisfies the equations of motion at every $n$-th order without the need to consider perturbative corrections if one imposes a specific condition on $\alpha\sim 1/N^{3/2}_{av}$, where $N_{av}=\sqrt{N_fN_i}$ is the geometric average of the total number of excitations (bosons) between the initial $\ket{N_i}$ and final $\ket{N_f}$ Fock states that the interaction is acting on, as shown in Appendix (A). At $n=0$ the excitations (bosons) are massless with energy $|E_k|=|\B{k}|$ and for any arbitrary $n>0$ a mass-like term is introduced through the self interaction that turns out to be the main driver of a $Q\cdot Q$ quadrupole-quadrupole type interaction.
This, alongside the imposed condition on $\alpha$, allows us to include all terms up to an arbitrary $n$-th order.

The EFT has to be suitable for describing nuclei, and therefore it is necessary to use discretized fields instead of Eq.\,\eqref{PW}, through localized plane waves within cubic elements of volume $V$ with periodic boundary conditions. This condition requires that the plane waves  have the following discrete form:
\begin{gather}
\varphi(\B{r},t)=\frac{1}{\sqrt{V}}\Big(\sum_{\B{k}}\frac{b^+_{\B{k}}}{\sqrt{|2E_k|}}e^{\iota k^{\mu}x_{\mu}} 
+ \frac{b^-_{\B{k}}}{\sqrt{|2E_k|}}e^{-\iota k^{\mu}x_{\mu}}\Big),
\label{df1}
\end{gather}
where $b^+_{\B{k}}$ creates an excitation (boson) and $b_{\B{k}}$ destroys an excitation (boson), respectively, with energy $|E_k|=|\B{k}|$ by acting on a $\ket{N_k}$ state, where $N_k$ is the number of excitations (bosons) with momentum $\B{k}$.

Now that we have identified the required fields and operators, we need for them to describe excitations of a system of $\mathcal{A}$ nucleons, which means the fields must enter pair-wise (quadratically, to preserve the parity of each single-nucleon wave function), and therefore for a nucleus with $\mathcal{A}$ nucleons the Lagrangian density given in Eq.\,\eqref{EFTLagrangian} has to be generalized to 
\begin{equation}
\mathcal{L}^{(n)}=\frac{\alpha^n}{2^{n+1}(n+1)!}\big(\partial_{\mu}\varphi_{p}\partial^{\mu}\varphi_{p}-nm^2\varphi_{p}^2\big)^{n+1}, \label{EFTLp}
\end{equation}
which is the Lagrangian density of an $\mathcal{A}$-component real scalar field and is O$(\mathcal{A}-1)$ symmetric [$(\mathcal{A}-1)$ to remove the center-of-mass contribution]. It has been established that the symplectic Sp(3,\B{R}) group is a complementary dual of the O$(\mathcal{A}-1)$ symmetry group [32]. This implies that the Lagrangian itself is part symplectic, meaning that the resulting quantum mechanical Hamiltonian from it, after making specific couplings, preserves symplectic symmetry and doesn’t mix configurations belonging to different symplectic irreps. As for the p subscript, it denotes the sum over all nucleons in the system

\subsection{The Sp(3,\B{R}) algebra and Symplectic basis}
The symplectic symmetry is the natural extension of the SU(3) symmetry and is realized by its 21 many-body generators in their Cartesian form. Since we are constructing a 4-dimensional EFT it is appropriate to represent the symplectic generators in the interaction picture (Heisenberg representation) where the operators explicitly depend on time. This is done through
\begin{equation}
b^{\pm}(t)=b^{\pm}e^{\pm \iota\Omega t}.
\end{equation}
Using this definition we get the following 
\begin{gather}
\nonumber
A_{ij}= \frac{1}{2}b^+_{ip}b^+_{jp}e^{2\iota\Omega t}, \\
B_{ij}= \frac{1}{2}b^-_{ip}b^-_{jp}e^{-2\iota\Omega t}, \label{SpO} \\
\nonumber
C_{ij}= \frac{1}{2}(b^+_{ip}b^-_{jp}+b^-_{jp}b^+_{ip}),
\end{gather}
where the $i,j$ in subscripts denote the spatial directions and the repeated index $p$ implies a sum over the number of nucleons in the system being described. The objects $2\mathcal{Q}_{ij}=C_{ij}+C_{ji}$ are the generators of the Elliott SU(3) group that act within a major harmonic oscillator shell, whereas the symplectic raising $A_{ij}$ operator and its conjugate $B_{ij}$ lowering operator connect states differing in energy by $2\Omega$, twice the harmonic oscillator energy. 
The interaction picture clearly states that the symplectic operators $A$ and $B$ are the ones responsible for the dynamics (they depend on time) in nuclei that can be interpreted as vibrations in space and time. Whereas the $C$ operators (independent of time) are responsible for the static deformed configurations in nuclei that can rotate freely. This was perhaps implicitly evident from the fact that the Sp(3,\B{R}) symmetry ($A$ and $B$) is the dynamical extension of the SU(3) symmetry ($C$), but now it is explicitly evident through their representation in the interaction picture.

To further understand the significance of these operators it is useful to define the following set of operators which are more suitable for a physical interpretation of the symplectic operators 
\begin{gather}
\nonumber
Q_{ij}= \mathcal{Q}_{ij}+A_{ij}+B_{ij}, \\
K_{ij}= \mathcal{Q}_{ij}-A_{ij}-B_{ij}, \label{PSPO}  \\
\nonumber
L_{ij}= -i(C_{ij}-C_{ji}), \\
\nonumber
S_{ij}= 2\iota(A_{ij}-B_{ji}), 
\end{gather}
where $Q_{ij}$ is the quadrupole tensor and is responsible for deformation, $K_{ij}$ is the the many-body kinetic tensor, $L_{ij}$ is the angular momentum tensor responsible for rotations and $S_{ij}$ is the vorticity tensor responsible for the flow of deformation.

Symplectic basis states are constructed by acting with the symplectic raising operator $A$ on the so-called band-head of the symplectic irrep $\ket{\sigma}$ which, is unique and is defined as $B\ket{\sigma}=0$ to be a lowest weight state. The band-head $\ket{\sigma}$ is mathematically similar to how the vacuum behaves $b\ket{0}=0$. However, unlike vacuum, $\ket{\sigma}$ can contain physical particles; namely, nucleons.

The complete labeling of a sympletic basis state that is constructed from its $\ket{\sigma}$ bandhead irrep is $\ket{\sigma n\rho\omega\kappa LM}$ where $\sigma\equiv N_{\sigma}(\lambda_{\sigma}\mu_{\sigma})$ labels the bandhead, $n\equiv N_n(\lambda_n\mu_n)$ labels the excited state that, coupled to the bandhead, yields the final configuration labeled by $\omega\equiv N_{\omega}(\lambda_{\omega}\mu_{\omega})$ with $\rho$ multiplicity, $L$ angular momentum with $\kappa$ multiplicity and its $M$ projection.  $N_{\omega}=N_{\sigma}+N_{n}$ is the total number of bosons (oscillator quanta), $\lambda_a=(N_a)_z-(N_a)_x$ and $\mu_a=(N_a)_x-(N_a)_y$ are the SU(3) quantum numbers, where $a=\sigma,\omega,n$. They denote the intrinsic deformation of the state since they count the difference in the number of oscillator quanta in the z and x, and x and y directions respectively.
These states are ideal for describing collective features of nuclei and for serving as basis states for the EFT. Similar to the Fock state notation $\ket{N_k}$, symplectic basis states also describe bosons numbered by $N_{\omega}$ making them a suitable basis state to which an application of our EFT interaction yields a quantum mechanical Hamiltonian that can be utilized for carrying out nuclear structure studies.

\subsection{SpEFT Hamiltonian}

The $n$-th order Hamiltonian density, derived in Appendix (B) is
{\allowdisplaybreaks
\begin{gather}
\mathcal{H}^{(n)}=\frac{\alpha^n}{2^{n+1}(n+1)!}\big(\dot{\varphi}_{p_1}^2-\varphi_{p_1}^{\prime}\cdot \varphi_{p_1}^{\prime}-nm^2\varphi_{p_1}^2\big)^n  \nonumber \\ \label{EFTHamiltonian}
\times\big((2n+1)\dot{\varphi}_{p_2}^2+\varphi_{p_2}^{\prime}\cdot \varphi_{p_2}^{\prime} + nm^2\varphi_{p_2}^2\big),
\end{gather}
where $\dot{\varphi}\equiv \frac{\partial\varphi}{\partial t}$ and $\varphi^{\prime}\equiv \frac{\partial\varphi}{\partial \B{r}}$.
The total Hamiltonian density at the $n$-th order is a sum over all possible $n+1$ combinations of the $n+1$-th term in the second parenthesis in Eq.\,\eqref{EFTHamiltonian} with respect to the $n$ terms in the first paranthesis, therefore it is Hermitian (see Appendix (B) for proof). However, for purposes of calculating matrix elements it is sufficient to only consider Eq.\,\eqref{EFTHamiltonian} and then add all the possible other combinatorial terms as described.
}

The coupled fields in Eq.\,\eqref{EFTHamiltonian} are 
\begin{gather}
\varphi_{p}^2=\frac{1}{V}\sum_{\B{k},\B{q}}\frac{1}{\sqrt{|2E_k||2E_q|}} \nonumber
\times \\ \nonumber
\Big(b^+_{p\B{k}}e^{\iota k^{\mu}x_{\mu}} + b^-_{p\B{k}}e^{-\iota k^{\mu}x_{\mu}}\Big) \Big(b^+_{p\B{q}}e^{\iota q^{\mu}x_{\mu}}+b^-_{p\B{q}}e^{-\iota q^{\mu}x_{\mu}} \Big), \\ \nonumber
\dot{\varphi}_{p}^2=\frac{1}{V}\sum_{\B{k},\B{q}}\frac{\iota E_k }{\sqrt{|2E_k|}}\frac{\iota E_q}{\sqrt{|2E_q|}} \times  \\ \nonumber 
\Big(b^+_{p\B{k}}e^{\iota k^{\mu}x_{\mu}} 
- b^-_{p\B{k}}e^{-\iota k^{\mu}x_{\mu}}\Big) \Big(b^+_{p\B{q}}e^{\iota q^{\mu}x_{\mu}}-b^-_{p\B{q}}e^{-\iota q^{\mu}x_{\mu}} \Big) , \\ \nonumber
\varphi_{p}^{\prime}\cdot\varphi_{p}^{\prime}=\frac{1}{V}\sum_{\B{k},\B{q}}\frac{(-\iota \B{k})}{\sqrt{|2E_k|}}\frac{(-\iota \B{q})}{\sqrt{|2E_q|}} \times  \\ \label{FP}
\Big(b^+_{p\B{k}}e^{\iota k^{\mu}x_{\mu}} - b^-_{p\B{k}}e^{-\iota k^{\mu}x_{\mu}}\Big) \Big(b^+_{p\B{q}}e^{\iota q^{\mu}x_{\mu}}-b^-_{p\B{q}}e^{-\iota q^{\mu}x_{\mu}} \Big) . 
\end{gather}
For convenience, from here on we will drop the index $p$ denoting the sum over particle numbers from the fields since they don't affect any of the follow-on derivations and can be recovered as may be required at any time. As evident from the formulas above the creation and annihilation operators enter into the Hamiltonian density in pairs of $b^+b^+$, $b^-b^-$, $b^+b^-$ and $b^-b^+$. 
This allows us to describe them through the symplectic operators defined in Eqs.\,\eqref{SpO}.
This definition further enables us to transition from $\ket{N_k}$ to $\ket{\sigma}$ 
where $N_{\omega}$ will be equivalent to a state with number of bosons $N_k$ created(destroyed) by $b^+_{\B{k}}(b^-_\B{k})$ with momentum $\B{k}$. 
Knowing this we can rewrite the fields in Eqs.\,\eqref{FP} as follows:
\begin{gather}
\varphi^2=\frac{1}{V}\sum_{\B{k},\B{q}}\frac{1}{\sqrt{|2E_k||2E_q|}} \nonumber
\times \\ \nonumber
\Big(Z^+_{\B{k}}Z^+_{\B{q}}+Z^-_{\B{k}}Z^-_{\B{q}}+Z^+_{\B{k}}Z^-_{\B{q}}+Z^-_{\B{k}}Z^+_{\B{q}} \Big), \\ \nonumber
\dot{\varphi}^2=\frac{1}{V}\sum_{\B{k},\B{q}}\frac{\iota E_k }{\sqrt{|2E_k|}}\frac{\iota E_q}{\sqrt{|2E_q|}} \times  \\ \nonumber 
\Big(Z^+_{\B{k}}Z^+_{\B{q}}+Z^-_{\B{k}}Z^-_{\B{q}}-Z^+_{\B{k}}Z^-_{\B{q}}-Z^-_{\B{k}}Z^+_{\B{q}} \Big) , \\ \nonumber
\varphi^{\prime}\cdot\varphi^{\prime}=\frac{1}{V}\sum_{\B{k},\B{q}}\frac{(-\iota\B{k})}{\sqrt{|2E_k|}}\frac{(-\iota\B{q})}{\sqrt{|2E_q|}} \times  \\ \label{ZFP}
\Big(Z^+_{\B{k}}Z^+_{\B{q}}+Z^-_{\B{k}}Z^-_{\B{q}}-Z^+_{\B{k}}Z^-_{\B{q}}-Z^-_{\B{k}}Z^+_{\B{q}} \Big) , 
\end{gather}
where, for further convenience we use the notation
$Z^{\pm}_{\B{k}}=b^{\pm}_{\B{k}}e^{\pm ik^{\mu}x_{\mu}}$. And finally, with these further simplifying definitions in play, the Hamiltonian density in Eq.\,\eqref{EFTHamiltonian} can be rewritten as follows:
\begin{gather}
 \label{Hn}
\mathcal{H}^{(n)}=\frac{1}{2^{n+1}(n+1)!}\frac{\alpha^n}{V^{n+1}}\sum_{\B{k}_1\B{k}_2...\B{k}_{n+1}}\sum_{\B{q}_1\B{q}_2...\B{q}_{n+1}} \\ \nonumber  \frac{\mathcal{Z}_1\mathcal{Z}_2......\mathcal{Z}_n\Xi_{n+1}}{2^{n+1}\sqrt{E_{k_1}E_{k_2}.....E_{k_n+1}E_{q_1}E_{q_2}.....E_{q_n+1}}}, 
\end{gather}

where we further use the following notation:
\begin{gather}
\mathcal{Z}_n=\bigg((-E_{k_n}E_{q_n}+\B{k}_n\cdot\B{q}_n-nm^2)(Z^+_{\B{k}_n}Z^+_{\B{q}_n}+Z^-_{\B{k}_n}Z^-_{\B{q}_n})\nonumber \\ -(-E_{k_n}E_{q_n}+\B{k}_n\cdot\B{q}_n+nm^2)(Z^+_{\B{k}_n}Z^-_{\B{q}_n}+Z^-_{\B{k}_n}Z^+_{\B{q}_n})\bigg).
\end{gather}

\begin{gather}
\Xi_{n+1}=\bigg((-(2n+1)E_{k_{n+1}}E_{q_{n+1}}-\B{k}_{n+1}\cdot\B{q}_{n+1}+nm^2) \nonumber \times \\ (Z^+_{\B{k}_{n+1}}Z^+_{\B{q}_{n+1}}+Z^-_{\B{k}_{n+1}}Z^-_{\B{q}_{n+1}}) \nonumber \\ -(-(2n+1)E_{k_{n+1}}E_{q_{n+1}}-\B{k}_{n+1}\cdot\B{q}_{n+1}-nm^2) \nonumber \times \\ (Z^+_{\B{k}_{n+1}}Z^-_{\B{q}_{n+1}}+Z^-_{\B{k}_{n+1}}Z^+_{\B{q}_{n+1}})\bigg).
\end{gather} 
With all of this in place, we finally come to an expression for the Hamiltonian density, $\mathcal{H}$, that enters into the integral for $H$ that we consider next.

\subsection{Diagonal coupling and Monopole Hamiltonian}
The Hamiltonian is 
\begin{equation}
H= \int_{-L}^{+L}\mathcal{H}dV.
\end{equation}
Substituting $\mathcal{H}$ into the integration we notice that for every order $n$ we have $n+1$ pairs of $Z^+Z^+$, $Z^+Z^-$ and their conjugates multiplied with each other inside the integration. To demonstrate this, let us consider a term like $Z^+Z^+$ for example for the simplest case $n=0$ which gives a term like
\begin{equation}
\int_{-L}^{+L}b_{\B{k}}^+(t)b_{\B{q}}^+(t)e^{ i(\B{k}+\B{q})\cdot\B{r}}dV,
\end{equation}
where we absorbed the time component of the exponent into the operators. This integral is zero because of the periodic  boundary condition on $\B{k}$ and $\B{q}$  unless $\B{k}+\B{q}=0$. To generalize for $n+1$ pairs, each term $Z^+Z^+$ and $Z^-Z^-$
has to have $\B{k}+\B{q}=0$, and each term $Z^+Z^-$ and $Z^-Z^+$ has to have $\B{k}-\B{q}=0$. We call this ``diagonal coupling'' because for each pair in the integral we couple $\B{k}$ to $\B{q}$ for example $\B{k}_1$ to $\B{q}_1$, $\B{k}_2$ to $\B{q}_2$ and $\B{k}_n$ to $\B{q}_n$, etc. The result of this simple coupling is that each $\mathcal{Z}$ term in Eq.\,\eqref{Hn} doesn't interact with other similar terms, meaning the sums don't mix with each other inside the integral, which leads to the Hamiltonian presented in Eq.\,\eqref{3-1}, below. (See Appendix (C) for this derivation,)
\begin{widetext}
\begin{gather}
H^{(n)}=\frac{1}{2^{n+1}(n+1)!}\sum_{\B{k}_1\B{k}_2...\B{k}_{n+1}}\frac{1}{2^{n+1}E_{k_1}E_{k_2}.....E_{k_n+1}}\frac{\alpha^n}{V^{n}}\times \nonumber \\ \bigg((-2E_{k_1}^2-2(2n-1)E_{k_1}^2)(b^+_{\B{k}_1}b^+_{-\B{k}_1}e^{2iE_{k_1}t}+b^-_{\B{k}_1}b^-_{-\B{k}_1}e^{-2iE_{k_1}t})\bigg)\times \nonumber \\ ..... \times
\bigg((-2E_{k_n}^2-2(2n-1)E_{k_n}^2)(b^+_{\B{k}_n}b^+_{-\B{k}_n}e^{2iE_{k_1}t}+b^-_{\B{k}_n}b^-_{-\B{k}_n}e^{-2iE_{k_1}t})\bigg)\times \nonumber \\
\bigg(-2E_{k_{n+1}}^2(b^+_{\B{k}_{n+1}}b^+_{-\B{k}_{n+1}}e^{2iE_{k_{n+1}}t}+b^-_{\B{k}_{n+1}}b^-_{-\B{k}_{n+1}}e^{-2iE_{k_{n+1}}t}) +(2n+2)E_{k_{n+1}}^2(b^+_{\B{k}_{n+1}}b^-_{\B{k}_{n+1}}+b^-_{\B{k}_{n+1}}b^+_{\B{k}_{n+1}})\bigg). \label{3-1}
\end{gather}
\end{widetext}
Comparing the terms in the Hamiltonian to the symplectic operators defined in Eq.\,\eqref{SpO} it is straightforward to see that this Hamiltonian is symplectic in nature because, for example, $b^-_{\B{k}_1}b^-_{-\B{k}_1}$ is simply $b^-_ib^-_i$ which is the symplectic lowering operator $B_{ii}$ after we recover the sum over particle number $p$.  This implies that symplectic symmetry emerges naturally from the EFT Lagrangian in Eq.\,\eqref{EFTLagrangian} whose sole construction was done naturally by extending the harmonic oscillator Lagrangian out to its $n$-th order. This implies that symplectic symmetry is an extension of the SU(3) symmetry of the harmonic oscillator which algebraically was known and well understood through nuclear physics applications but here, as seen through these developments, symplectic symmetry in nuclear physics has its origin at a more fundamental level than previously considered; specifically, it derives from and is underpinned by a logical EFT formulation.

In this effective field theory for determining the structure of atomic nuclei the nucleons in a nucleus are represented through the total energy quanta (bosons) they bring forward, which in turn can be created and destroyed at all possible energy values. This can be envisioned as having an infinite quantum mechanical harmonic oscillator systems, each with $E_k$ wherein the nucleons are contained. Such excitations are constrained to a very narrow range of possible energy values.  Studies done with mean field models and also realistic interactions all support such a claim, as do calculations using the NCSpM \cite{PhysRevC.89.034312} and SA-NCSM \cite{PhysRevLett.124.042501}. Specifically, in the latter cases, one typically finds that utilizing a single symplectic irrep suffices to recover nearly $70\%$-$80\%$ of the probability distribution, and more specifically, accounts for nearly $90\%$-$100\%$ of observables, such as energy spectra, B(E2) values and radii.

Given the fact that in all such cases the excitation quanta include only a single energy mode; $\hbar\Omega=41\mathcal{A}^{-1/3}$ MeV,
the application of the Hamiltonian on a single symplectic state further reduces it to only one term, where from each sum over $\B{k}$ survives; namely, the term where $E_k=\hbar\Omega$ which reduces Eq.\, \eqref{3-1} to
\begin{gather}
H^{(n>1)}_d=(-n)^n(2\hbar\Omega)^{n+1}\frac{\alpha^n}{V^n}(A_{ii}+B_{ii})^n\times \nonumber \\
\big((n+1)C_{jj}-(A_{jj}+B_{jj})\big). \label{3-2}
\end{gather}
This is a quantum mechanical Hamiltonian that is the natural extension to the harmonic oscillator Hamiltonian for $n\geq 1$. It represents a one-body interaction extended to an arbitrary $n$-th order. This results from the diagonal coupling discussed above and hence the $d$ subscript, and is solely responsible for generating monopole excitations in nuclei that do not contribute to the dynamics since they are just powers of $A_{ii}+B_{ii}$. They destroy the leading order harmonic oscillator at every $n\geq 1$ which is unphysical and hence they have to be removed from the final Hamiltonian. What is responsible for dynamics are vibrations in space and time due to the quadrupole excitations in nuclei that result from off-diagonal couplings in the Hamiltonian in Eq.\,\eqref{Hn}. The only term from the diagonal coupling that contributes to the Hamiltonian is the harmonic oscillator which, is $H^{(0)}=\hbar\Omega C_{ii}$.

\subsection{Off-Diagonal coupling the Quadrupole Hamiltonian}
In this section, we consider two pairs of Z's, $Z^+Z^+$, for example inside the integral resulting from multiplying two $\mathcal{Z}$ in Eq.\,\eqref{Hn}, namely, 
\begin{equation}
\int_{-L}^{+L}b_{\B{k}_1}^+(t)b_{\B{q}_1}^+(t)b_{\B{k}_2}^+(t)b_{\B{q}_2}^+(t)e^{ i(\B{k}_1+\B{q}_1+\B{k}_2+\B{q}_2)\cdot\B{r}}dV.
\end{equation}
For $n=1$, which as we discussed before, this will be zero unless $\B{k}_1+\B{q}_1+\B{k}_2+\B{q}_2=0$. We managed this before by picking $\B{k}_1+\B{q}_1=0$ and $\B{k}_2+\B{q}_2=0$, etc., which resulted to the diagonal coupling. However there are many possibilities to make $\B{k}_1+\B{q}_1+\B{k}_2+\B{q}_2=0$. What is particularly interesting is if we choose $\B{k}_1+\B{q}_2=0$ and $\B{k}_2+\B{q}_1=0$. This results in a pair of $A_{\B{k}_1-\B{k}_2}A_{\B{k}_2-\B{k}_1}$ which creates a boson pair with momentum $\B{k}_1$ and $-\B{k}_2$, respectively, and creates another pair with momentum $\B{k}_2$ and $-\B{k}_1$, respectively, such that the total momentum of both pairs is conserved. If we pick $\B{k}_1=-\B{k}_2$ this will result to the diagonal coupling Hamiltonian in Eq.\, \eqref{3-2} derived in the previous section. However we can pick $|\B{k}_1|=|\B{k}_2|$ such that $\B{k}_1\perp\B{k}_2$. This reduces $A_{\B{k}_1-\B{k}_2}A_{\B{k}_2-\B{k}_1}$ to $A_{ij}A_{ji}$ which creates two boson pairs in the $i$-th and $j$-th direction such that $i\neq j$. The same argument applies to other terms like $Z^+Z^+Z^-Z^-$, $Z^+Z^+Z^+Z^-$, $Z^+Z^+Z^+Z^-$, $Z^-Z^-Z^+Z^-$ etc. 

The expression for the off-diagonal Hamiltonian depends on $n$. If $n$ is odd then we have $n+1$ even pairs of $Z^{\pm}$ that all could be coupled to each other resulting in $(n-1)/2$ identical pairs and one unique pair.
If $n$ is even then we have $n+1$ odd pairs, from which we can form either $n/2$ identical pairs, and a unique pair or $(n-2)/2$ identical pairs and two unique pairs.
This results in three off-diagonal Hamiltonians, one for odd $n$ and two for even $n$ for $n>0$, since for $n=0$ only diagonal coupling is possible. The resulting expressions will contain terms like $A_{ij}A_{ji}$, $C_{ij}C_{ji}$, $B_{ij}C_{ji}$ etc., which can be represented in terms of $Q_{ij}$ and $K_{ij}$ resulting into the following three two-body Hamiltonian expansions, see Appendix (D) for this derivation. The expansion resulting from the off-diagonal coupling in Eq.\,\eqref{Hn} at every $n=odd$ is
\begin{gather}
H^{(n=odd)}_{od}=\frac{(\hbar\Omega)^{n+1}}{2^{n+1}}\frac{\alpha^n}{V^{n}}\times \nonumber \\ \big(g_n^2Q_{ij}Q_{ji}+K_{ij}K_{ji} -g_n\lbrace Q_{ij},K_{ji}\rbrace\big)^{(n-1)/2} \times \nonumber \\\big(-g_n^2Q_{ij}Q_{ji}+(2n+1)K_{ij}K_{ji} -ng_n\lbrace Q_{ij},K_{ji}\rbrace\big). 
\end{gather}
In the above equation, $g_n$ denotes the strength of the quadrupole operator and is tied to the mass parameter introduced in Eq.\,\eqref{EFTLagrangian} (see Appendix (A) for derivation).
The expansion resulting from the off-diagonal coupling in Eq.\,\eqref{Hn} at every $n=even$ is
\begin{gather}
H^{(n=even)}_{od}=\frac{(\hbar\Omega)^{n+1}}{2^{n+1}}\frac{\alpha^n}{V^{n}}\times \nonumber \\ \big(g_n^2Q_{ij}Q_{ji}+K_{ij}K_{ji}-g_n\lbrace Q_{ij},K_{ji}\rbrace\big)^{n/2} \times \nonumber \\ \big((n+1)C_{ll}-(A_{jj}+B_{jj})\big). 
\end{gather}
This is one of the expansions resulting from the off-diagonal coupling in Eq.\,\eqref{Hn} at every $n=even$ and the other one is removed since every term is proportional to $A_{ii}+B_{ii}$, see Appendix (D) for additional details.

\section{Analysis and Results}

Incorporating all the above considerations, the final quantum mechanical Hamiltonian is as follows:
\begin{equation}
H= \hbar\Omega C_{ii} + \sum_{n=1}H^{(n=odd)}_{od}+\sum_{n=2}H^{(n=even)}_{od}. \label{QMH}
\end{equation}
Each $n$-th term in this Hamiltonian is a ($n+1$)-body interaction. Therefore this Hamiltonian, as formulated below, includes all possible interaction terms up to infinity, but excludes terms resulting from triple, quadruple or even higher off-diagonal couplings that are naturally associated with new power series of 3-body and 4-body character, for example terms like $Q_{ij}Q_{jf}Q_{fi}$ and $Q_{ij}Q_{jf}Q_{fl}Q_{li}$, respectively. These terms lie outside the scope of the present paper as here we have chosen to limit the theory to at most two off-diagonal coupling terms, and the resulting interactions and their respective powers.

In this section we will outline in subsection (A) how the parameters of the theory are chosen, how the effective parameters of the Hamiltonian tie to the parameters introduced in the Lagrangian density, and finally discuss their physical implications. In subsection (B) the dynamical effects of the interaction and time average of the Hamiltonian will be derived. Finally in subsection (C) we will present some result for applications of this Hamiltonian to ${}^{20}$Ne, ${}^{22}$Ne and ${}^{22}$Mg.

\subsection{Parameters of the EFT}


The resulting Hamiltonian for this EFT [Eq.\eqref{QMH}] is effectively a two parameter theory; the parameters being $\frac{\alpha}{V}\hbar\Omega$ and $g_n$. The parameter $\frac{\alpha}{V}\hbar\Omega=\frac{V_{\mathcal{A}}}{b^3N^3_{av}}$ establishes a clear seperation of scales. It is simply the ratio of the average spherical nuclear volume $V_A=\frac{4}{3}\pi R^3$, where $R=1.2\mathcal{A}^{1/3}$ is the average radius of the ground state, over the volume corresponding to the harmonic oscillator excitations $V=b^3N_{av}^{3/2}$, where $b$ is the oscillator length, along with the plane-wave condition, $\alpha\sim 1/N^{3/2}_{av}$ that allows further stretching of $V$.  


If $\frac{V_{\mathcal{A}}}{b^3N^3_{av}}<< 1$ the average volume of the nucleus is less than a volume determined by adding up the total number of its excitation quanta, when a plane wave solution is valid therefore preserving the harmonic oscillator structure. But if $\frac{V_{\mathcal{A}}}{b^3N^3_{av}} \sim 1$, signaling that the volume determined by adding up the number of oscillator excitations in play is approaching the average volume, a plane wave solution becomes untenable with higher order corrections becoming ever more relevant, ending in a complete breakdown when the $Q\cdot Q$ interactions destroys the HO structure. In short, as more nucleons are added to the system their corresponding boson excitations have to also increase appropriately such that they can be represented by plane waves as indicated by the $\frac{V_{\mathcal{A}}}{b^3N^3_{av}}$ measure. This further emphasizes the fact that this scale is valid as long as a shell structure description is appropriate for the systems being studied.

The second parameter of the theory is $g_n$ which is the strength of the quadrupole operator because only the quadrupole terms in Eq.\,\eqref{QMH} carry this parameter as a multiplier. The value of this parameter determines the strength with which the quadrupole tensor enters into any analysis relative to that of the kinetic tensor. In particular, it should be clear that if $g_n=0$ one gets a  power series in $K_{ij}K_{ji}$. It is therefore important that the value chosen for $g_n>1$ for $\forall n$ is required to balance the interaction between $Q_{ij}Q_{ji}$, $K_{ij}K_{ji}$ and $Q_{ij}K_{ji}$.

The parameter $g_n$ is expressed in terms of the mass-like parameter introduced in Eq.\,\eqref{massterm} to represent the strength of the interaction. It results from off-diagonal couplings that depend on $n$ which has a simple physical interpretation,
\begin{equation}
g_n=\frac{2n-1}{n}g.
\end{equation}
The consequences of this is that the energy of the bosons contributing to the formation of a given final symplectic configuration, starting from an initial one, increases uniformly as more pairs are considered. This simple picture suggests that if $g_1=g$ and $g_{\infty}=2g$ then as $n\to\infty$ the weight of $Q_{ij}Q_{ji}$ will scale as
\begin{equation}
\lim_{n\to\infty}\frac{(\hbar\Omega)^{n+1}}{2^{n+1}}\frac{\alpha^n}{V^{n}}g_n^{n+1}=0.
\end{equation}
Although this means there will be a new parameter $g_n$ at every order of the Hamiltonian, they are all determined once $g$ is fixed, and therefore the theory is effectively a two parameter theory; namely, $\frac{V_A}{b^3N^3_{av}}$ and $g$.
In applications of the theory, unlike $\frac{V_A}{b^3N^3_{av}}$, $g$ can be fitted to known observables.


\subsection{Time Average of the Hamiltonian}

The Hamiltonian in Eq.\,\eqref{QMH} depends on time implicitly. This dependence comes through the symplectic operators defined in Eqs.\,\eqref{SpO} and since they enter in pairs, for example, in the two-body interaction terms they will have the following time factors $e^{\pm4\iota\Omega t}$ for $A_{ij}A_{ji}$ and $B_{ij}B_{ji}$, $e^{\pm2\iota\Omega t}$ for $A_{ij}C_{ji}$ and $B_{ij}C_{ji}$, unity for $A_{ij}B_{ji}$ and $C_{ij}C_{ji}$ (``$+$ for $A$s and ``$-$'' for $B$s). It is evident that the time independent terms, like $A_{ij}B_{ji}$  and $C_{ij}C_{ji}$ are responsible for rotations. As for the dynamical terms, like $A_{ij}A_{ji}$  and $B_{ij}B_{ji}$, they are responsible for vibrations in nuclei.

The time dependence has to be integrated out. This is done by averaging the Hamiltonian over the time period that the nucleons interact with each other.
\begin{equation}
H=\frac{1}{T}\int^T_0 H(t) dt,
\end{equation}
where $H(t)$ is the Hamiltonian given in Eq.\,\eqref{QMH} with its time dependence written explicitly. $T$ is the upper time limit in which the strong interaction propagates and it is of order
$T=\frac{2R}{c}=\frac{2\times10^{-15}}{3\times10^8}\sim 10^{-23}s$. This allows us to evaluate the integral in the limit of $T\to0$ which implies that the self interacting fields in our EFT interact almost simultaneously
\begin{equation}
H=\lim_{T\to0}\frac{1}{T}\int^T_0 H(t) dt.
\end{equation}
The time-independent terms come out of this integral unchanged. As for the time-dependent ones, let us show an explicit example like $e^{4\iota\Omega t}$ which will be to the power of $\frac{n+1}{2}$ for $n=odd$ and to power of $\frac{n}{2}$ for $n=even$. For the odd ones we have
\begin{equation}
\lim_{T\to0}\frac{1}{T}\int^T_0 e^{2(n+1)\iota\Omega t} dt=\lim_{T\to0}\frac{e^{2(n+1)\iota\Omega T} -1}{2(n+1)\iota\Omega T}=1 .
\end{equation}
This proves that the time-dependent terms also come out unchanged except they drop their exponential time factors. The same proof applies to the even expansion terms as well.
So finally, the Hamiltonian in Eq.\,\eqref{QMH} could be applied to any nucleus as though it is independent of time.

\subsection{${}^{20}$Ne, ${}^{22}$Ne and ${}^{22}$Mg}

The ground state rotational bands for ${}^{20}$Ne, ${}^{22}$Ne and ${}^{22}$Mg all display a structure that is close to that of a rigid rotor, which makes them good tests of our EFT to see if the theory can reproduce this rotational behavior. Moreover, the $K_{ij}K_{ji}$ and $Q_{ij}K_{ji}$ in Eq.\,\eqref{QMH} should introduce irrotational-like departures from the simple $L(L+1)$ rule for the spectrum of a rigid rotor, which can as well be seen  in the ${}^{20}$Ne, ${}^{22}$Ne, and ${}^{22}$Mg set, making these ideal candidates for probing a range of rotational features within the context of our EFT theory.

\begin{figure} 
	\centering 
	\includegraphics[scale=0.27]{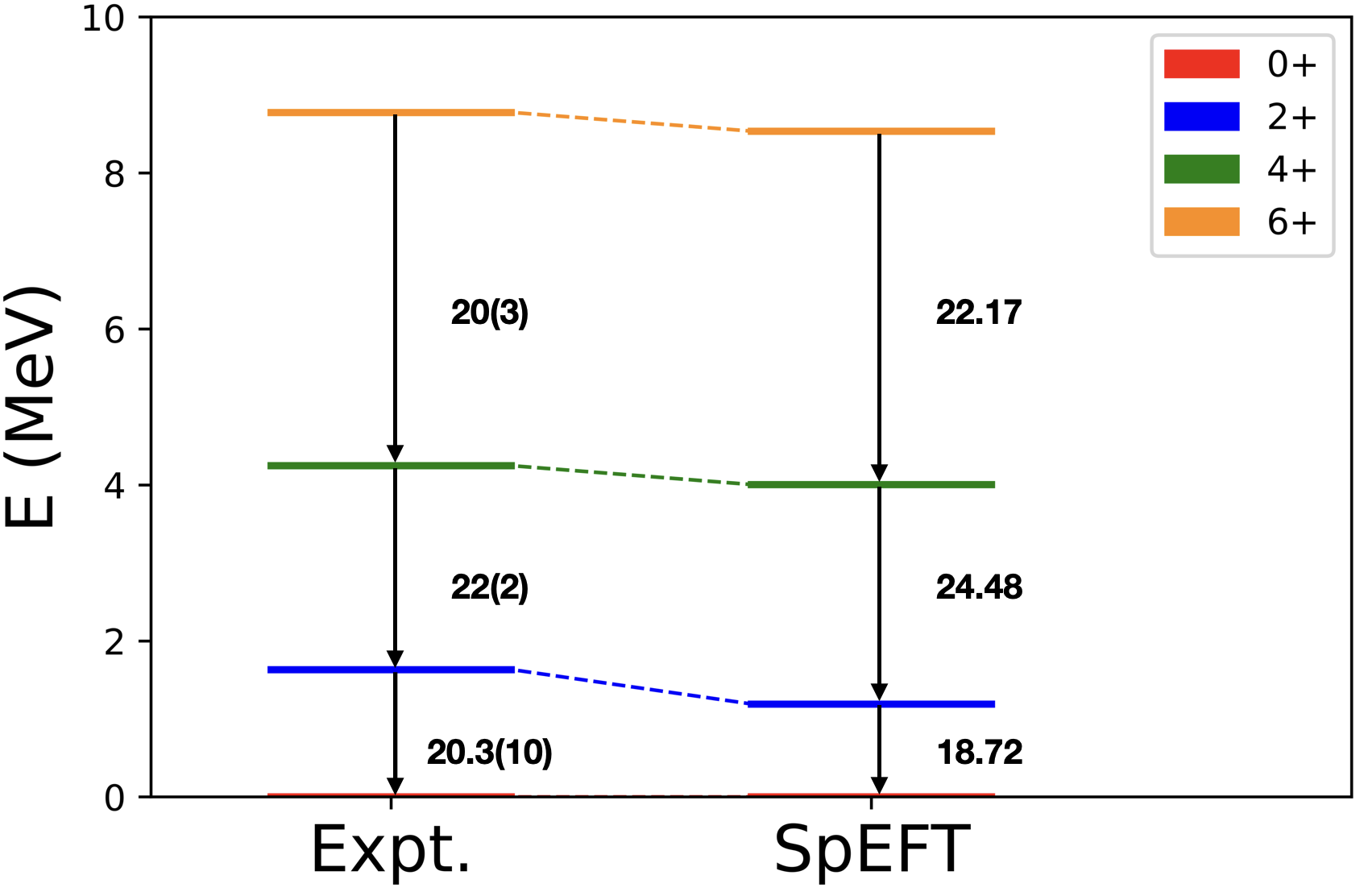}
	\caption{The energy spectrum and B(E2) values of the $48.5(8,0)$ symplectic irrep for ${}^{20}$Ne using the SpEFT in a $N_{\rm{max}}=12$ model space (EFT) compared to experimental data (Expt.) \cite{TILLEY1998249}. B(E2) values are in W.u.}
	\label{f1}
\end{figure}

\begin{figure} 
	\centering 
	\includegraphics[scale=0.27]{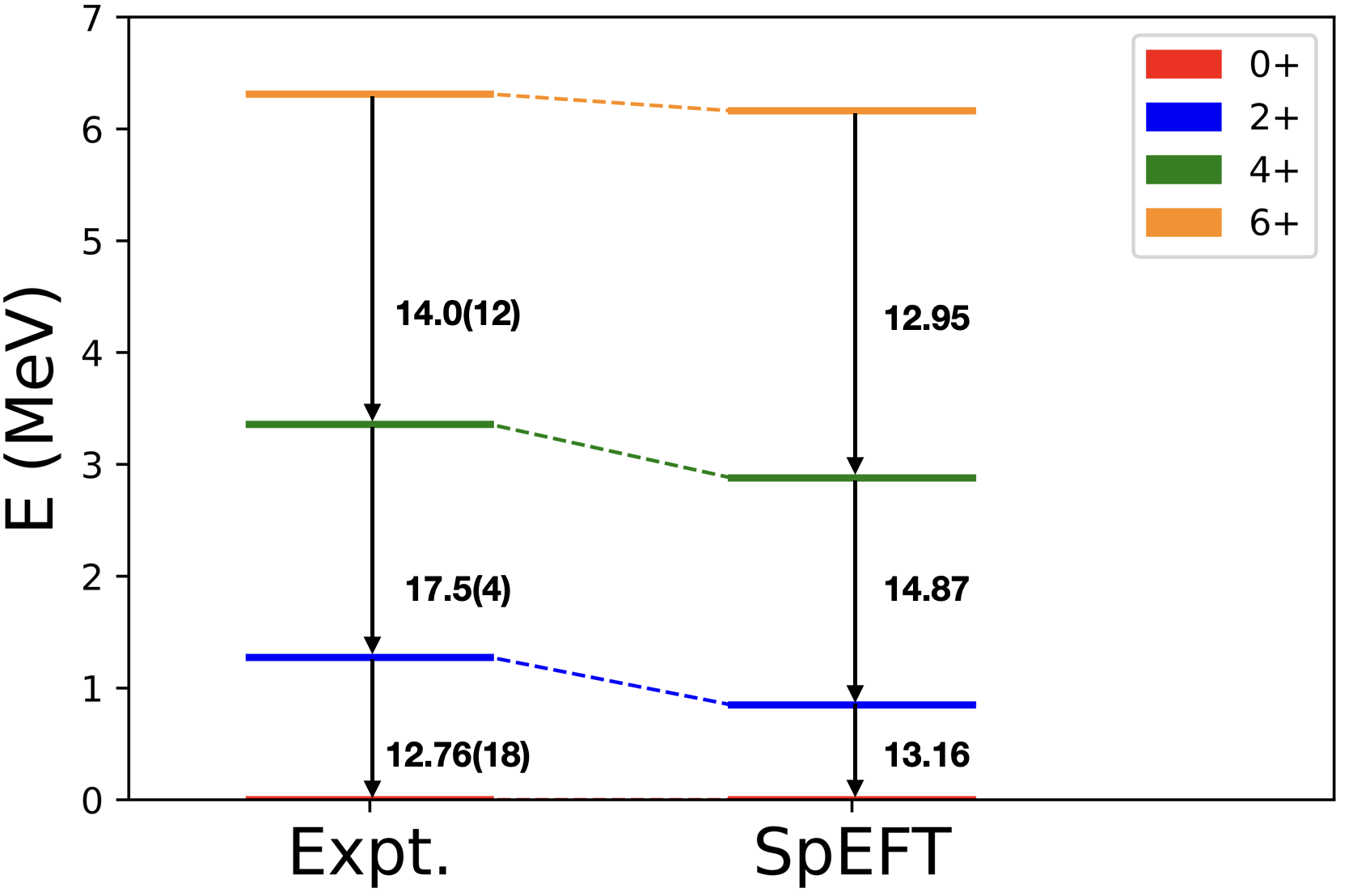}
	\caption{SpEFT energy spectrum and B(E2) values of the $55.5(8,2)$ symplectic irrep for ${}^{22}$Ne in a $N_{\rm{max}}=12$ model space compared to experimental data (Expt.) \cite{BASUNIA201569}. B(E2) values are in W.u. }
\label{f2}	
\end{figure}

\begin{figure} 
	\centering 
	\includegraphics[scale=0.27]{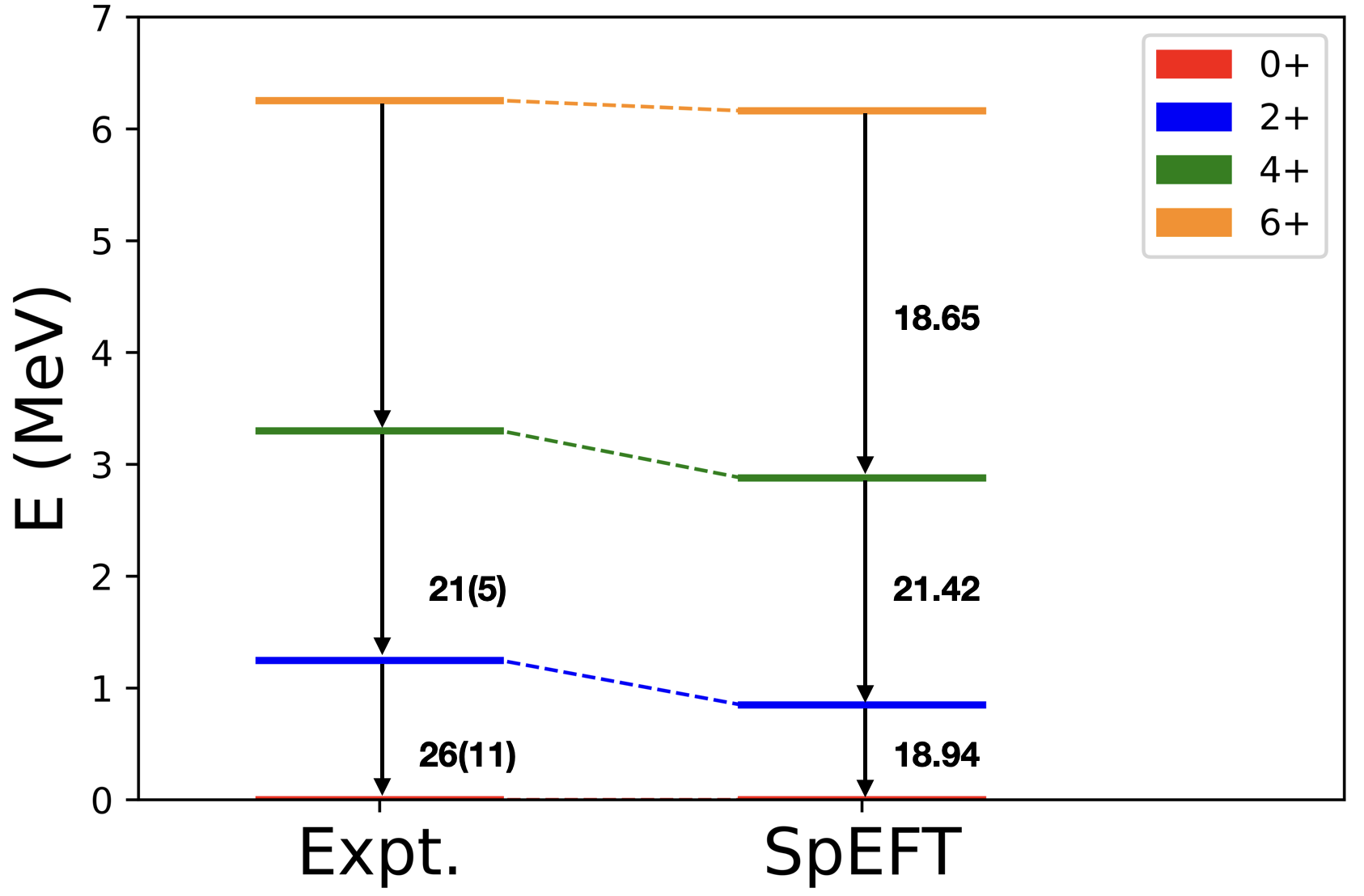}
	\caption{SpEFT energy spectrum and B(E2) values of the $55.5(8,2)$ symplectic irrep for ${}^{22}$Mg in a $N_{\rm{max}}=12$ model space compared to experimental data (Expt.) \cite{BASUNIA201569}. B(E2) values are in W.u. }
	\label{f3}
\end{figure}

Specifically, we carried out SpEFT calculations in the $N_{\rm{max}}=12$ model space, which was required for gaining good convergence of the observed B(E2) transitions, by adjusting only one parameter; $g=14$ for ${}^{20}$Ne and $g=14.7$ for ${}^{22}$Ne and ${}^{22}$Mg. For these nuclei, the  $N_{\rm{max}}=12$ model space is down-selected to only one spin-0 leading symplectic irrep; namely $48.5(8,0)$ for ${}^{20}$Ne, $55.5(8,2)$ for ${}^{22}$Ne and ${}^{22}$Mg. These selections also proved sufficient to simultaneously reproduce reasonably well-converged values for the observed energy spectra and nuclear radii. And most importantly, the results very clearly demonstrate that the SpEFT is able to do this without the need for introducing effective charges which is confirmed pictorially in Fig. (\ref{f1},\ref{f2},\ref{f3}), and by a comparison of the rms radii given in Table (\ref{tablerms}) that are as well in very good agreement with observations, all with only a single fitting parameter, $g$. 

In Table (\ref{tablescale}) we further give the maximum and minimum values of the scale parameter, $\frac{V_{\mathcal{A}}}{b^3N^3_{av}}$, for these nuclei.  All the presented results are calculated by including terms up to $n\leq 4$ in the Hamiltonian, which, as stressed above, was necessary for gaining good convergence of the theory to known observables. Additionally, we note that for the case of ${}^{20}$Ne, terms with $4<n\leq 6$ contribute $\sim -0.004$ MeV to the ground state energy and $\sim -0.004$ W.u. to the $2^+\rightarrow 0^+$ B(E2) transition. And beyond  these data-focused measures, it is interesting to note that these calculations were all carried out on a laptop, taking from about 10 minutes for the ${}^{20}$Ne case and up to approximately 2 hours for the ${}^{22}$Ne and ${}^{22}$Mg cases, a feature which serves to stress that as complex as the underpinning algebraic structure may seem to be (Section II together with the associated Appendices), its applications are computationally quite simple, even to the point of rendering the formalism suitable for more pedagogical uses.


\begin{table}
\caption{SpEFT matter rms radii $r_m$ (fm) of the ground states compared to their experimental counterparts for the nuclei under consideration \cite{CHULKOV1996219,SUZUKI1998661}.}
\centering \label{tablerms}
\begin{tabular}{ |m{1.5cm}|m{1.5cm}||m{1.5cm}| }
\hline
Nucleus & SpEFT & Expt.  \\
\hline
${}^{20}$Ne & 2.80 & 2.87(3)  \\
\hline
${}^{22}$Ne & 2.84 & -----   \\
\hline
${}^{22}$Mg & 2.84 & 2.89(6)  \\
\hline

\end{tabular}
\end{table}

\begin{table}
\caption{The maximum and minimum value of the scale of SpEFT, $\frac{V_{\mathcal{A}}}{b^3N^3_{av}}$ when $N_{av}=N_{\sigma}$, $N_{av}=N_{\sigma}+N_{max}$ accordingly.}
\centering \label{tablescale}
\begin{tabular}{ |m{1.5cm}|m{3cm}||m{3cm}| }
\hline
Nucleus & The maximum scale & The minimum scale \\
\hline
${}^{20}$Ne & $2.79\times 10^{-4}$ & $1.44\times 10^{-4}$  \\
\hline
${}^{22}$Ne & $1.95\times 10^{-4}$ & $1.08\times 10^{-4}$   \\
\hline
${}^{22}$Mg & $1.95\times 10^{-4}$ & $1.08\times 10^{-4}$  \\
\hline

\end{tabular}
\end{table}

\section{Conclusions}

In this paper we outlined a step-by-step process for how symplectic symmetry emerges from a simple self-interacting scalar field theory extended to $n$-th order. We then went on to identify the scale of this interaction and imposed the necessary conditions that justify treating the scalar fields as plane waves. Furthermore, the application of the fields onto a single symplectic irrep generated a quantum mechanical Hamiltonian. It is composed of a harmonic oscillator at leading order and a dominant quadrupole-quadrupole interaction at next-to-leading, and higher orders. These results explain why, from an EFT perspective, phenomenological models using a simple harmonic oscillator with quadrupole type interactions have been successful in capturing the relevant physics in nuclei. Their origin lies in a simple scalar-field theory framework. Moreover, for the first time, we identified the dynamical operators of the symplectic algebra, and showed how they explicitly behave under vibrations in time.

The resulting SpEFT Hamiltonian is a complex, yet simple interaction to study nuclear observables. It can produce energy spectra, enhanced electromagnetic transitions, and rms matter radii without the need for effective charges. Moreover, it does this with only a single fitted parameter $g$. The main advantage of this theory lies in its simplicity to explain how deformation arises and drives nuclear dynamics. This key feature allows the SpEFT to take advantage of the underlying symmetry, and therefore doesn't require access to large-scale computational resources. This was successfully demonstrated by its application to ${}^{20}$Ne, ${}^{22}$Ne, ${}^{22}$Mg and its ability to produce results that are in very reasonable agreement with experiment on just a laptop.

\begin{acknowledgments}
%
Work supported by: DK \& JPD -- Louisiana State University (College of Science, Department of Physics \& Astronomy) as well as the Southeastern Universities Research Association (SURA, a U.S. based Non-profit Organization);
VIM -- In part by the U.S. Department of Energy (SURA operates the Thomas Jefferson National Accelerator Facility for the U.S.;  Department of Energy under Contract No. DE-AC05-06OR23177); and CDR -- National Natural Science Foundation of China (Grant No.\,12135007).
\end{acknowledgments}

\appendix

\section{Plane Wave Solution}
Let us find the condition on $\alpha$ for which the plane wave solution given by Eq.\,\eqref{PW} satisfies the equations of motion for the Lagrangian density in Eq.\,\eqref{EFTLp}.
At a specific $n$ order we will have
\begin{equation}\label{E-Ln}
\partial^{\mu}\frac{\partial \mathcal{L}^{(n)}}{\partial\varphi_{\pi,\mu}}=\frac{\partial \mathcal{L}^{(n)}}{\partial\varphi_{\pi}}.
\end{equation}
Calculating each derivative for the $\pi$-th field yields
\begin{equation}
\frac{\partial \mathcal{L}^{(n)}}{\partial\varphi_{\pi}}=\alpha^n\frac{n+1}{2^{n}(n+1)!}\big(\partial_{\mu}\varphi_p\partial^{\mu}\varphi_p-nm^2\varphi^2_p\big)^{n}(-nm^2\varphi_{\pi}),
\end{equation}
\begin{equation} 
\frac{\partial \mathcal{L}^{(n)}}{\partial\varphi_{\pi,\mu}}=\alpha^n\frac{n+1}{2^{n}(n+1)!}\big(\partial_{\nu}\varphi_p\partial^{\nu}\varphi_p-nm^2\varphi^2_p\big)^{n}\partial_{\mu}\varphi_{\pi}, \label{A3}
\end{equation}
\begin{gather}
\partial^{\mu}\frac{\partial \mathcal{L}^{(n)}}{\partial\varphi_{\pi,\mu}}=\alpha^n\frac{n+1}{2^{n}(n+1)!}\big(\partial_{\eta}\varphi_p\partial^{\eta}\varphi_p-nm^2\varphi^2_p\big)^{n}\partial^{\mu}\partial_{\mu}\varphi_{\pi}  \nonumber \\ +
\alpha^n\frac{n(n+1)}{2^{n}(n+1)!}\big(\partial_{\eta}\varphi_p\partial^{\eta}\varphi_p-nm^2\varphi^2_p\big)^{n-1}\times \nonumber \\ \big(2\partial_{\nu}\partial^{\nu}\varphi_{\pi}\partial^{\mu}\varphi_p\partial_{\mu}\varphi_p-2n m^2\varphi_p\partial^{\mu}\varphi_p\partial_{\mu}\varphi_{\pi}\big).
\end{gather}
Now summing over all possible $n$ we get the equation of motion 
\begin{gather}
\sum_{n=0}^{\infty}\alpha^n\frac{n+1}{2^{n}(n+1)!}\big(\partial_{\eta}\varphi_p\partial^{\eta}\varphi_p-nm^2\varphi^2-p\big)^{n} \times \nonumber \\ (\partial^{\mu}\partial_{\mu}\varphi_{\pi} +nm^2\varphi_{\pi}) \nonumber \\ +
\alpha^n\frac{n(n+1)}{2^{n}(n+1)!}\big(\partial_{\eta}\varphi_p\partial^{\eta}\varphi_p-nm^2\varphi^2_p\big)^{n-1}\times \nonumber \\ \big(2\partial_{\nu}\partial^{\nu}\varphi_{\pi}\partial^{\mu}\varphi_p\partial_{\mu}\varphi_p-2n m^2\varphi_p\partial^{\mu}\varphi_p\partial_{\mu}\varphi_{\pi}\big)=0.
\label{eqofmotion}
\end{gather}
Since the total Lagrangian density is $\mathcal{L}=\sum_n\mathcal{L}^{(n)}$ there is a sum over $n$. For $n=0$ we will have $\partial^{\mu}\partial_{\mu}\varphi_{\pi}=0$ which is the Klein-Gordon equation for massless bosons and the plane wave solution given in Eq.\,\eqref{PW} satisfies it with $E^2=k^2$ therefore the first $n=0$ term disappears. As for $n>0$, by plugging $\partial^{\mu}\partial_{\mu}\varphi_{\pi}=0$ into Eq.\,\eqref{eqofmotion} and since $\alpha^n\frac{n+1}{2^{n}(n+1)!}\big(\partial_{\eta}\varphi_p\partial^{\eta}\varphi_p-nm^2\varphi^2_p\big)\neq 0$ otherwise the Lagrangian density would be zero so we can take it out of the equation. For $\forall n>0$ term we must have 
\begin{gather}
\alpha^n\Big((\partial_{\eta}\varphi_p\partial^{\eta}\varphi_p-nm^2\varphi^2_p)(nm^2\varphi_{\pi}) \nonumber \\ -n(2n m^2\varphi_p\partial^{\mu}\varphi_p\partial_{\mu}\varphi_{\pi})\Big)=0.
\label{sumeq}
\end{gather}
The above equation is not zero because the plain wave solution doesn't satisfy it. However, if we pick the magnitude of $\alpha$ correctly for the leading order ($n=1$) then the above equation can be made to be much less than one, and close to zero. Since $\varphi_p\sim (b^+_p + b^-_p)$ then the average maximum magnitude of  $\bra{N_f}\varphi^2_p\ket{N_i}\sim N_{av}$ where $N_{av}=\sqrt{N_fN_i}$ is the geometric average of the total number of excitations (bosons) between the initial and final Fock states. Since the above equation of motion is proportional to $\varphi_{\pi}\varphi^2_p$ and the minimum upper limit of $\bra{f}\varphi^2_p\varphi_{\pi}\ket{i}\sim N_{av}\sqrt{N_{av}}$ then for $n=1$ we roughly have
\begin{equation}
\alpha (N_{av}\sqrt{N_{av}}) \sim 0.
\end{equation}
Therefore, we have the following estimate for the magnitude of $\alpha$
\begin{equation}
\alpha\sim \frac{1}{N_{av}\sqrt{N_{av}}}.
\label{Alphaorder}
\end{equation}
Establishing this magnitude also allows us to estimate the magnitude of $m^2$. To do this we need the field derivatives for which we have
\begin{gather}
\partial^{\mu}\varphi=\frac{1}{(2\pi)^{3/2}}\int_{-\infty}^{+\infty}\iota k_{\mu}\psi(\B{k},E)e^{\iota k^{\nu}x_{\nu}}dEd\B{k}, \nonumber \\
\partial_{\mu}\varphi\partial^{\mu}\varphi=-\frac{1}{(2\pi)^{3}}\int_{-\infty}^{+\infty}\int_{-\infty}^{+\infty}k_{1\mu}k_2^{\mu}\psi(\B{k}_1,E)\psi(\B{k}_2,E) \times \nonumber \\ e^{\iota (k_1^{\nu}+k_2^{\nu})x_{\nu}}dE_1d\B{k}_1dE_2d\B{k}_2.
\end{gather}
Using these and plugging them into Eq.\,\eqref{sumeq} we get the following parametric equation (keeping in mind that for purposes of applying this to nuclei we would only keep one specific value of $E_{\B{k}}=\hbar\Omega$ and hence the integrals drop out): 
\begin{gather}
\sum_{n=1}^{\infty}\big(-|\B{k}_1||\B{k}_2|+\B{k}_1\cdot\B{k}_2-nm^2\big)nm^2 \nonumber \\ +2n^2m^2\big(|\B{k}_1||\B{k}_2|-\B{k}_1\cdot\B{k}_2\big)=0.
\end{gather}
Which gives the following formula for the mass parameter
\begin{equation}
m^2=\frac{2n-1}{n}\big(|\B{k}_1||\B{k}_2|-\B{k}_1\cdot\B{k}_2\big).
\end{equation}
Therefore this choice of the parameter guarantees that the plane wave satisfies the equation of motion for a specific energy value $|\B{k}_1|=|\B{k}_2|=\sqrt{\frac{g}{n}}\hbar\Omega$. This choice further reduces the formula to the following three cases
\begin{equation}
nm^2=\begin{cases} g_n\hbar\Omega^2, & \mbox{if $\B{k}_1 \bot \B{k}_2$} \\ 2\hbar\Omega^2, & \mbox{if $\B{k}_1$ is anti-parallel to $\B{k}_2$} \\ 0. & \mbox{if $\B{k}_1$ is parallel to $\B{k}_2$}
\end{cases}
\label{massterm}
\end{equation}
In the above equation where $g_n=\frac{2n-1}{n}g$ was introduced only for the $\B{k}_1 \bot \B{k}_2$ case since it is an effective description to the weight of the resulting interaction from such coupling between the fields, where $g$ captures its strength.   Given that $\alpha$ is determined by Eq.\,\eqref{Alphaorder}, $g$ is the only parameter that will be fitted from one nuclear system to the other.

\section{Hamiltonian derivation and Hermiticity}
The Hamiltonian at any $n$ order is given by the Legendre transformation
\begin{equation}\label{B1}
\mathcal{H}^{(n)}=\dot{\varphi}\frac{\partial\mathcal{L}^{(n)}}{\partial\dot{\varphi}}-\mathcal{L}^{(n)}
\end{equation}
Using the Eq.\,\eqref{A3} and plugging $\mu=0$ we get the following
\begin{equation}\label{B2}
\frac{\partial \mathcal{L}^{(n)}}{\partial\dot{\varphi}}=\alpha^n\frac{n+1}{2^{n+1}(n+1)!}\big(\partial_{\nu}\varphi\partial^{\nu}\varphi-nm^2\varphi^2\big)^{n}2\dot{\varphi}.
\end{equation}
Plugging it back in the Hamiltonian formula we get
\begin{widetext}
\begin{gather}
\mathcal{H}^{(n)}=\alpha^n\frac{2(n+1)}{2^{n+1}(n+1)!}\big(\partial_{\nu}\varphi\partial^{\nu}\varphi-nm^2\varphi^2\big)^{n}\dot{\varphi}^2 - \frac{\alpha^n}{2^{n+1}(n+1)!}\big(\partial_{\mu}\varphi\partial^{\mu}\varphi-nm^2\varphi^2\big)^{n+1},
\end{gather}
\end{widetext}
combining the terms we get
\begin{gather} 
\mathcal{H}^{(n)}=\frac{\alpha^n}{2^{n+1}(n+1)!}\big(\partial_{\nu}\varphi\partial^{\nu}\varphi-nm^2\varphi^2\big)^{n}\times \nonumber \\ 
\big((2n+1)\dot{\varphi}^2+\varphi^{\prime}\cdot\varphi^{\prime}+nm^2\varphi^2\big). \label{B4}
\end{gather}
The Hamiltonian in Eq.\,\eqref{B4} is not Hermitian 
but the Hamiltonian in Eq.\,\eqref{B1} is by definition Hermitian. This is due the fact when we calculated the term $\frac{\partial \mathcal{L}^{(n)}}{\partial\dot{\varphi}}$ we neglected all possible other combinations of $\dot{\varphi}$ with $\big(\partial_{\nu}\varphi\partial^{\nu}\varphi-nm^2\varphi^2\big)$. In reality this term should be
\begin{widetext}
\begin{gather}
\frac{\partial \mathcal{L}^{(n)}}{\partial\dot{\varphi}}=\alpha^n\frac{2(n+1)}{2^{n+1}(n+1)!}\Bigg(\dot{\varphi}\big(\partial_{\nu}\varphi\partial^{\nu}\varphi-nm^2\varphi^2\big)^{n} + \big(\partial_{\nu}\varphi\partial^{\nu}\varphi-nm^2\varphi^2\big)\dot{\varphi}\big(\partial_{\nu}\varphi\partial^{\nu}\varphi-nm^2\varphi^2\big)^{n-1} \nonumber \\ + \big(\partial_{\nu}\varphi\partial^{\nu}\varphi-nm^2\varphi^2\big)^2\dot{\varphi}\big(\partial_{\nu}\varphi\partial^{\nu}\varphi-nm^2\varphi^2\big)^{n-2} + .......... + \big(\partial_{\nu}\varphi\partial^{\nu}\varphi-nm^2\varphi^2\big)^{n}\dot{\varphi}\Bigg). \label{B5}
\end{gather}
\end{widetext}
As evident from Eq.\,\eqref{B5} there are $n+1$ terms which are different combinations of $\dot{\varphi}$ with $\big(\partial_{\nu}\varphi\partial^{\nu}\varphi-nm^2\varphi^2\big)$. This results into the following Hamiltonian
\begin{widetext}
\begin{gather} 
\mathcal{H}^{(n)}=\frac{\alpha^n}{2^{n+1}(n+1)!}\Bigg(\big((2n+1)\dot{\varphi}^2+\varphi^{\prime}\cdot\varphi^{\prime}+nm^2\varphi^2\big)\big(\partial_{\nu}\varphi\partial^{\nu}\varphi-nm^2\varphi^2\big)^{n} \nonumber \\ + \big(\partial_{\nu}\varphi\partial^{\nu}\varphi-nm^2\varphi^2\big)\big((2n+1)\dot{\varphi}^2+\varphi^{\prime}\cdot\varphi^{\prime}+nm^2\varphi^2\big)\big(\partial_{\nu}\varphi\partial^{\nu}\varphi-nm^2\varphi^2\big)^{n-1} \nonumber \\ 
+ \big(\partial_{\nu}\varphi\partial^{\nu}\varphi-nm^2\varphi^2\big)\big((2n+1)\dot{\varphi}^2+\varphi^{\prime}\cdot\varphi^{\prime}+nm^2\varphi^2\big)\big(\partial_{\nu}\varphi\partial^{\nu}\varphi-nm^2\varphi^2\big)^{n-2} \nonumber \\
+ ...... + \big(\partial_{\nu}\varphi\partial^{\nu}\varphi-nm^2\varphi^2\big)^{n}  \big((2n+1)\dot{\varphi}^2+\varphi^{\prime}\cdot\varphi^{\prime}+nm^2\varphi^2\big)\Bigg). \label{B6}
\end{gather}
\end{widetext}
The above Hamiltonian is Hermitian. Since the formula for the Hamiltonian is very long we refrained from writing down all the terms and limited ourselves to only the $n+1$-th term 
because it is easier for purposes of deriving a quantum mechanical Hamiltonian applicable for describing nuclear phenomena.
Note that we also refrained from writing down all possible $n!$ combinations of $\big(\partial_{\nu}\varphi\partial^{\nu}\varphi-nm^2\varphi^2\big)$ with itself since they are all identical for matrix element calculations hence the normalization factor of $(n+1)!$. $n!$ for identical combinations of $\big(\partial_{\nu}\varphi\partial^{\nu}\varphi-nm^2\varphi^2\big)$ combined with $n+1$ combinations with the $\big((2n+1)\dot{\varphi}^2+\varphi^{\prime}\cdot\varphi^{\prime}+nm^2\varphi^2\big)$ term as we saw above.

\section{Diagonal coupling calculation}
As discussed in section 3, we will pick $\B{k}=-\B{q}$ for $Z^+Z^+$ and $Z^-Z^-$ and pick $\B{k}=\B{q}$ for $Z^+Z^-$ and $Z^-Z^+$ so they will survive the integration over the volume which results to
\begin{widetext}
\begin{gather}
H^{(n)}=\frac{1}{2^{n+1}(n+1)!}\sum_{\B{k}_1\B{k}_2...\B{k}_{n+1}}\frac{1}{2^{n+1}E_{k_1}E_{k_2}.....E_{k_n+1}}\frac{\alpha^n}{V^{n}}\times \nonumber \\ \bigg((-E_{k_1}^2-\B{k}_1^2-nm^2)(b^+_{\B{k}_1}b^+_{-\B{k}_1}e^{2\iota E_{k_1}t}+b^-_{\B{k}_1}b^-_{-\B{k}_1}e^{-2\iota E_{k_1}t})-(-E_{k_1}^2+\B{k}_1^2+nm^2)(b^+_{\B{k}_1}b^-_{\B{k}_1}+b^-_{\B{k}_1}b^+_{\B{k}_1})\bigg)\times \nonumber \\
\bigg((-E_{k_2}^2-\B{k}_2^2-nm^2)(b^+_{\B{k}_2}b^+_{-\B{k}_2}e^{2\iota E_{k_2}t}+b^-_{\B{k}_2}b^-_{-\B{k}_2}e^{-2\iota E_{k_2}t})-(-E_{k_2}^2+\B{k}_2^2+nm^2)(b^+_{\B{k}_2}b^-_{\B{k}_2}+b^-_{\B{k}_2}b^+_{\B{k}_2})\bigg)\times \nonumber \\
....................................................... \times \nonumber \\
\bigg((-E_{k_n}^2-\B{k}_n^2-nm^2)(b^+_{\B{k}_n}b^+_{-\B{k}_n}e^{2\iota E_{k_n}t}+b^-_{\B{k}_n}b^-_{-\B{k}_n}e^{-2\iota E_{k_n}t})-(-E_{k_n}^2+\B{k}_n^2+nm^2)(b^+_{\B{k}_n}b^-_{\B{k}_n}+b^-_{\B{k}_n}b^+_{\B{k}_n})\bigg)\times \nonumber \\
\bigg((-(2n+1)E_{k_{n+1}}^2+\B{k}_{n+1}^2+nm^2)(b^+_{\B{k}_{n+1}}b^+_{-\B{k}_{n+1}}e^{2\iota E_{k_{n+1}}t}+b^-_{\B{k}_{n+1}}b^-_{-\B{k}_{n+1}}e^{-2\iota E_{k_{n+1}}t}) \nonumber \\ -(-(2n+1)E_{k_{n+1}}^2-\B{k}_{n+1}^2-nm^2)(b^+_{\B{k}_{n+1}}b^-_{\B{k}_{n+1}}+b^-_{\B{k}_{n+1}}b^+_{\B{k}_{n+1}})\bigg). \label{D1}
\end{gather}
\end{widetext}
But $E_{k_1}^2=\B{k}_1^2$. As for $m^2$ we have $m^2=\frac{2n-1}{n}(2E^2_{k})$ for $\B{k}=-\B{q}$ and $m^2=0$ for $\B{k}=\B{q}$   by definition which further reduces Eq.\,\eqref{D1} to 
\begin{widetext}
\begin{gather}
H^{(n)}=\frac{1}{2^{n+1}(n+1)!}\sum_{\B{k}_1\B{k}_2...\B{k}_{n+1}}\frac{1}{2^{n+1}E_{k_1}E_{k_2}.....E_{k_n+1}}\frac{\alpha^n}{V^{n}}\times \nonumber \\ \bigg((-2E_{k_1}^2-2(2n-1)E_{k_1}^2)(b^+_{\B{k}_1}b^+_{-\B{k}_1}e^{2\iota E_{k_1}t}+b^-_{\B{k}_1}b^-_{-\B{k}_1}e^{-2\iota E_{k_1}t})\bigg)\times \nonumber \\ ..... \times
\bigg((-2E_{k_n}^2-2(2n-1)E_{k_n}^2)(b^+_{\B{k}_n}b^+_{-\B{k}_n}e^{2\iota E_{k_1}t}+b^-_{\B{k}_n}b^-_{-\B{k}_n}e^{-2\iota E_{k_1}t})\bigg)\times \nonumber \\
\bigg(-2E_{k_{n+1}}^2(b^+_{\B{k}_{n+1}}b^+_{-\B{k}_{n+1}}e^{2\iota E_{k_{n+1}}t}+b^-_{\B{k}_{n+1}}b^-_{-\B{k}_{n+1}}e^{-2\iota E_{k_{n+1}}t}) +(2n+2)E_{k_{n+1}}^2(b^+_{\B{k}_{n+1}}b^-_{\B{k}_{n+1}}+b^-_{\B{k}_{n+1}}b^+_{\B{k}_{n+1}})\bigg). \label{D2}
\end{gather}
\end{widetext}
Its application on a single symplectic state further reduces this Hamiltonian. Only one term from each sum over $\B{k}$ will survive namely the term where $E_k=\hbar\Omega$ which gives us
\begin{gather}
H^{(n)}=\frac{(-1)^n}{2^{n+1}(n+1)!}\frac{1}{2^{n+1}(\hbar\Omega)^{n+1}}\frac{\alpha^n}{V^{n}}\times \nonumber \\ \bigg(4n\hbar\Omega^2(b^+_{\B{k}_1}b^+_{-\B{k}_1}e^{2\iota\Omega t}+b^-_{\B{k}_1}b^-_{-\B{k}_1}e^{-2\iota\Omega t})\bigg)\times \nonumber \\ \bigg(4n\hbar\Omega^2(b^+_{\B{k}_2}b^+_{-\B{k}_2}e^{2\iota\Omega t}+b^-_{\B{k}_2}b^-_{-\B{k}_2}e^{-2\iota\Omega t})\bigg)\times \nonumber \\
..... \times
\bigg(4n\hbar\Omega^2(b^+_{\B{k}_n}b^+_{-\B{k}_n}e^{2i\iota\Omega t}+b^-_{\B{k}_n}b^-_{-\B{k}_n}e^{-2\iota\Omega t})\bigg)\times \nonumber \\
\bigg(-2\hbar\Omega^2(b^+_{\B{k}_{n+1}}b^+_{-\B{k}_{n+1}}e^{2\iota\Omega t}+b^-_{\B{k}_{n+1}}b^-_{-\B{k}_{n+1}}e^{-2\iota\Omega t}) \nonumber \\ +(2n+2)\hbar\Omega^2(b^+_{\B{k}_{n+1}}b^-_{\B{k}_{n+1}}+b^-_{\B{k}_{n+1}}b^+_{\B{k}_{n+1}})\bigg). \label{D3}
\end{gather}
Now using the definitions in Eqs.\,\eqref{SpO} and recovering the particle numbers we get 
\begin{gather}
H^{(n)}=\frac{(-n)^n}{2^{n+1}(n+1)!}\frac{4^{n+1}(\hbar\Omega)^{2(n+1)}}{2^{n+1}(\hbar\Omega)^{n+1}}\frac{\alpha^n}{V^n}(2A_{ii}+2B_{ii})^n\times \nonumber \\
\big(-2(A_{jj}+B_{jj}) +(2n+2)C_{jj}\big). \label{D4}
\end{gather}
Carrying out the necessary reductions we will finally get
\begin{gather}
H^{(n)}=(-n)^n(2\hbar\Omega)^{n+1}\frac{\alpha^n}{V^n}(A_{ii}+B_{ii})^n\times \nonumber \\
\big(-(A_{jj}+B_{jj}) +(n+1)C_{jj}\big). \label{D5}
\end{gather}
Note that we accounted for all possible combinations of the terms in respect to exchanging places with each other therefore the factorial term in front reduces to unity as discussed earlier in Appendix B. 

\section{Off-diagonal coupling calculation}
The off-diagonal coupling will result to the following possible couplings for $\forall n$ presented in Table (\ref{table}), of all possible terms in pairs to survive the integration over volume of the Hamiltonian density in Eq.\,\eqref{Hn}.
\begin{table}[ht]
\caption{All possible couplings in pairs of two Z operators. There are 16 possible terms. Only 8 are shown since the other 8 are conjugates and have identical couplings}
\centering \label{table}
\begin{tabular}{ |m{2.8cm}|m{4cm}| }
\hline
$Z^+_{\B{k}_n}Z^+_{\B{q}_n}Z^+_{\B{k}_{n+1}}Z^+_{\B{q}_{n+1}}$ & $\B{k}_n=-\B{q}_{n+1}$,$\B{k}_{n+1}=-\B{q}_n$   \\
\hline
$Z^+_{\B{k}_n}Z^+_{\B{q}_n}Z^-_{\B{k}_{n+1}}Z^-_{\B{q}_{n+1}}$ & $\B{k}_n=\B{q}_{n+1}$,$\B{k}_{n+1}=\B{q}_n$   \\
\hline
$Z^+_{\B{k}_n}Z^+_{\B{q}_n}Z^+_{\B{k}_{n+1}}Z^-_{\B{q}_{n+1}}$ & $\B{k}_n=\B{q}_{n+1}$,$\B{k}_{n+1}=-\B{q}_n$   \\
\hline
$Z^+_{\B{k}_n}Z^+_{\B{q}_n}Z^-_{\B{k}_{n+1}}Z^+_{\B{q}_{n+1}}$ & $\B{k}_n=-\B{q}_{n+1}$,$\B{k}_{n+1}=\B{q}_n$   \\
\hline
$Z^+_{\B{k}_n}Z^-_{\B{q}_n}Z^+_{\B{k}_{n+1}}Z^-_{\B{q}_{n+1}}$ & $\B{k}_n=\B{q}_{n+1}$,$\B{k}_{n+1}=\B{q}_n$   \\
\hline
$Z^+_{\B{k}_n}Z^-_{\B{q}_n}Z^-_{\B{k}_{n+1}}Z^+_{\B{q}_{n+1}}$ & $\B{k}_n=-\B{q}_{n+1}$,$\B{k}_{n+1}=-\B{q}_n$   \\
\hline
$Z^-_{\B{k}_n}Z^+_{\B{q}_n}Z^-_{\B{k}_{n+1}}Z^+_{\B{q}_{n+1}}$ & $\B{k}_n=\B{q}_{n+1}$,$\B{k}_{n+1}=\B{q}_n$   \\
\hline
\end{tabular}
\end{table}
To avoid diagonal terms like $A_{ii}$ $B_{ii}$ and $C_{ii}$ resulted from $\B{k}_n=\B{k}_{n+1}$ for $\forall n$ we pick $|\B{k}_n|=|\B{k}_{n+1}|$ such that $\B{k}_n\perp\B{k}_{n+1}$ for $\forall n$. This results into terms like $A_{ij}A_{ji}$ for example.
As discussed for any two $\mathcal{Z}$ terms in Eq. \eqref{Hn} we have
\begin{gather}
4(E^2+nm^2)^2(A_{ij}A_{ji}+B_{ij}B_{ji}+A_{ij}B_{ji}+B_{ij}A_{ji}) \nonumber \\ -4(E^4-n^2m^4)(A_{ij}\mathcal{Q}_{ji}+B_{ij}\mathcal{Q}_{ji}+\mathcal{Q}_{ij}B_{ji}+\mathcal{Q}_{ij}A_{ji}) \nonumber \\ +4(E^2-nm^2)^2\mathcal{Q}_{ij}\mathcal{Q}_{ji}. \label{E1}
\end{gather}
Expressing the symplectic operators in Eq. \eqref{E1} in terms of the quadrupole and kinetic tensors we get
\begin{gather}
(E^2+nm^2)^2(Q_{ij}Q_{ji}+K_{ij}K_{ji}-Q_{ij}K_{ji}-K_{ij}Q_{ji}) \nonumber \\ -(E^4-n^2m^4)(2Q_{ij}Q_{ji}-2K_{ij}K_{ji}) \nonumber \\ +(E^2-nm^2)^2(Q_{ij}Q_{ji}+K_{ij}K_{ji}+Q_{ij}K_{ji}+K_{ij}Q_{ji}) = \nonumber \\
4n^2m^4Q_{ij}Q_{ji}+4E^4K_{ij}K_{ji}-4nm^2E^2(Q_{ij}K_{ji}+K_{ij}Q_{ji}). \label{E2}
\end{gather}
As for the unique term resulting from coupling the $\mathcal{Z}_n$-th term to $\Xi_{n+1}$-th term in Eq.\,\eqref{Hn} we have
\begin{gather}
4((2n+1)E^4-n^2m^4+2n^2m^2E^2) \nonumber \\ \times(A_{ij}A_{ji}+B_{ij}B_{ji}+A_{ij}B_{ji}+B_{ij}A_{ji}) \nonumber \\ +4(-(2n+1)E^4-n^2m^4-2(n+1)nm^2E^2) \nonumber \\ \times(A_{ij}\mathcal{Q}_{ji}+B_{ij}\mathcal{Q}_{ji}) \nonumber \\ +4(-(2n+1)E^4-n^2m^4+2(n+1)nm^2E^2) \nonumber \\ \times(\mathcal{Q}_{ij}B_{ji}+\mathcal{Q}_{ij}A_{ji}) \nonumber \\ +4((2n+1)E^4-n^2m^4-2n^2m^2E^2)\mathcal{Q}_{ij}\mathcal{Q}_{ji}. \label{E3}
\end{gather}
Expressing the symplectic operators in Eq.\,\eqref{E3} in terms of the quadrupole and kinetic tensors we get
\begin{gather}
((2n+1)E^4-n^2m^4+2n^2m^2E^2) \nonumber \\\times(Q_{ij}Q_{ji}+K_{ij}K_{ji}-Q_{ij}K_{ji}-K_{ij}Q_{ji}) \nonumber \\ +(-(2n+1)E^4-n^2m^4-2(n+1)nm^2E^2) \nonumber \\ \times(Q_{ij}Q_{ji}-K_{ij}K_{ji}+Q_{ij}K_{ji}-K_{ij}Q_{ji}) \nonumber \\ +(-(2n+1)E^4-n^2m^4+2(n+1)nm^2E^2) \nonumber \\ \times(Q_{ij}Q_{ji}-K_{ij}K_{ji}-Q_{ij}K_{ji}+K_{ij}Q_{ji}) \nonumber \\ +((2n+1)E^4-n^2m^4-2n^2m^2E^2) \nonumber \\ \times(Q_{ij}Q_{ji}+K_{ij}K_{ji}+Q_{ij}K_{ji}+K_{ij}Q_{ji}). \label{E4}
\end{gather}
Finally combining them gives
\begin{gather}
-4n^2m^4Q_{ij}Q_{ji}+4(2n+1)E^4K_{ij}K_{ji} \nonumber \\-4n^2m^2E^2(Q_{ij}K_{ji}+K_{ij}Q_{ji}) \nonumber \\ -4(n+1)nm^2E^2(Q_{ij}K_{ji}-K_{ij}Q_{ji}).  \label{E5}
\end{gather}
Note that the last term in Eq.\,\eqref{E5} is not Hermitian. However as we explained in Appendix B there will be a Hermitian conjugate term resulting from other possible perturbations. For every possible term where we have $(Q_{ij}K_{ji}-K_{ij}Q_{ji})$ we will get its conjugate from other perturbations resulting to $-(Q_{ij}K_{ji}-K_{ij}Q_{ji})$ which cancel each other. Keeping this in mind we get the following for the Hamiltonian for an arbitrary odd $n$ where $E=\hbar\Omega$
\begin{gather}
H^{(n=odd)}_{od}=\frac{(n+1)!}{2^{n+1}(n+1)!}\frac{\hbar\Omega^{2(n+1)}}{(2\hbar\Omega)^{n+1}}\frac{\alpha^n}{V^{n}}\times \nonumber \\ \big(4g_n^2Q_{ij}Q_{ji}+4K_{ij}K_{ji}-4g_n\lbrace Q_{ij},K_{ji}\rbrace\big)^{(n-1)/2} \times \nonumber \\\big(-4g_n^2Q_{ij}Q_{ji}+4(2n+1)K_{ij}K_{ji} -4ng_n\lbrace Q_{ij},K_{ji}\rbrace\big), \label{E6}
\end{gather}
where $\lbrace Q_{ij},K_{ji}\rbrace$ is the anti-commutator and $g_n=nm^2/\hbar\Omega^2$. Canceling the coefficients we get 
\begin{gather}
H^{(n=odd)}_{od}=\frac{(\hbar\Omega)^{n+1}}{2^{n+1}}\frac{\alpha^n}{V^{n}}\times \nonumber \\ \big(g_n^2Q_{ij}Q_{ji}+K_{ij}K_{ji}-g_n\lbrace Q_{ij},K_{ji}\rbrace\big)^{(n-1)/2} \times \nonumber \\ \big(-g_n^2Q_{ij}Q_{ji}+(2n+1)K_{ij}K_{ji} -ng_n\lbrace Q_{ij},K_{ji}\rbrace\big). \label{E7}
\end{gather}
For $n>0$ and $n=even$ we have two options as discussed. The first option is to couple all identical terms off-diagonally and couple the $n+1$-th term diagonally which results to
\begin{gather}
H^{(n=even)}_{od}=\frac{(\hbar\Omega)^{n+1}}{2^{n+1}}\frac{\alpha^n}{V^{n}}\times \nonumber \\ \big(g_n^2Q_{ij}Q_{ji}+K_{ij}K_{ji}-g_n\lbrace Q_{ij},K_{ji}\rbrace\big)^{n/2} \times \nonumber \\\big((n+1)C_{ll}-(A_{ll}+B_{ll})\big). \label{E8}
\end{gather}
The other possibility is having one of the identical terms coupled diagonally and the rest off-diagonally which gives terms proportional to $(A_{ll}+B_{ll})$ and therefore we won't consider them either since they also change the size at every order.

\bibliographystyle{unsrt}
\bibliography{SFT}

\end{document}